# A new mechanism for gain in time dependent media

J.B. PENDRY[1,*], E. GALIFFI[1], AND P. A. HUIDOBRO[2]

[1]*The Blackett Laboratory, Department of Physics, Imperial College London, London, SW7 2AZ, UK .*
[2]*Instituto de Telecomunicações, Instituto Superior Tecnico-University of Lisbon, Avenida Rovisco Pais 1, Lisboa, 1049-001 Portugal.*

[*]j.pendry@imperial.ac.uk

**Abstract:** Time dependent systems do not in general conserve energy invalidating much of the theory developed for static systems and turning our intuition on its head. This is particularly acute in *luminal* space time crystals where the structure moves at or close to the velocity of light. Conventional Bloch wave theory no longer applies, energy grows exponentially with time, and a new perspective is required to understand the phenomenology. In this letter we identify a new mechanism for amplification: the compression of lines of force that are nevertheless conserved in number.

## 1. Introduction

Energy can be added to electromagnetic waves in several different fashions. We identify a mechanism, distinct from conventional ones, in which compression of lines of force is the active ingredient. There are instances of this in other contexts: a superconductor repels magnets because the magnetic lines of forces are compressed as they are rejected by the superconductor. A more dramatic example is the generation of thousand Tesla magnetic fields by explosive collapse of a copper cylinder enclosing magnetic lines of force [1]. Here we invoke the concept in the context of amplifying electromagnetic waves. We show that in some circumstances the number of lines of force, electric and magnetic, in a time dependent system is conserved and amplification occurs when these lines of force are squeezed closer together.

In this letter we use a simple model of a time dependent grating synthetically moving with a uniform velocity, $c_g = \Omega/g$,

$$\begin{aligned}\varepsilon(x-c_g t) &= \varepsilon_1 + 2\alpha_\varepsilon \cos(gx-\Omega t)\\ \mu(x-c_g t) &= \mu_1 + 2\alpha_\mu \cos(gx-\Omega t)\end{aligned} \quad (1)$$

where $g$ and $\Omega$ are spatial and temporal modulation frequencies, and $\alpha_\varepsilon, \alpha_\mu$ the strength of electric and magnetic modulations, respectively. The model has been extensively deployed in other studies of time dependent systems and a recent review is to be had in [2]. We assume that $\varepsilon, \mu$ are both real. Complications arising when the medium is dispersive and lacking in periodicity are discussed later in the paper. We stress that the medium itself does not move so that there is no restriction from relativity on the magnitude of $c_g$ which may take any value

$$0 < c_g < \infty .$$

Similar travelling-wave media have been investigated theoretically in the past [3, 4, 5, 6, 7]. More recently, they gained renewed interest thanks to their ability to break Lorentz reciprocity without need for an applied magnetic bias which can be exploited for use as isolators [8, 9, 10], as well as their topological [9], non-Hermitian [11] and cloaking [12] features. They have also recently been analysed and homogenised as effective media [13,14]. The luminal modulation regime considered here has recently been proposed for pulse formation [15,16] and broadband, nonreciprocal amplification [16].

However, a fundamental explanation of the physical mechanism responsible for this amplification has never been developed.

When waves interact with static structures we have many tools not only for calculating but also for understanding the processes and for conceptualizing a problem before we even begin to calculate. In periodic structures the Bloch wave vector is conserved and together with frequency, the other conserved quantity, gives a wealth of understanding. Its dispersion with frequency tells us where the band gaps are and where we are likely to find surface states. Bloch waves are the basis for understanding the topology of the states through such quantities as the Berry phase [17] and Chern number [18]. We stress that these concepts are not merely computational devices but central to our thought processes as tools for understanding and creating such systems.

In this paper we seek to provide a set of tools for understanding time dependent structures for which the traditional static concepts fail, by identifying a conserved quantity in the form of the number of lines of force contained in the system.

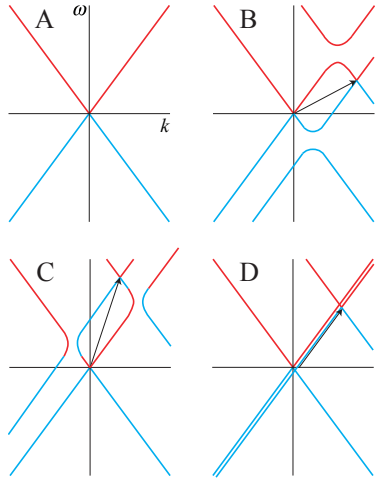

Fig. 1. A sketch of dispersion relation of light travelling through **A**: a uniform dielectric. **B**: a grating modulated as shown in Eq. (1) where $c_g < c_0/\sqrt{\varepsilon_1\mu_1}$. **C**: a grating modulated as shown in Eq. (1) where $c_g > c_0/\sqrt{\varepsilon_1\mu_1}$. **D**: the case of $c_g = c_0/\sqrt{\varepsilon_1\mu_1}$ where all the forward travelling states are degenerate and therefore strongly coupled. Red lines originate at positive frequencies, cyan lines at negative ones. The arrows show displacement of the bands by a space-time reciprocal lattice vector, $(g,\Omega)$, and their slope is the grating velocity

At first sight Eq. (1) would imply a straightforward generalization of Bloch's theorem mixing together waves differing by a space-time reciprocal lattice vector,

$$k',\omega' = k + ng, \omega + n\Omega \qquad (2)$$

so that as well as a Bloch wave vector, $k$, there is a Bloch frequency $\omega$, both conserved modulo $(g,\Omega)$. This is a good description of the problem for $c_g \ll c_0/\sqrt{\varepsilon_1 \mu_1}$ where $c_0/\sqrt{\varepsilon_1 \mu_1}$ is the velocity of light in the background medium. Fig. 1A shows dispersion of light in the background medium and Fig. 1B what happens when the grating is turned on: band gaps open in the normal way but now there is asymmetry about $k=0$ due to the breaking of time reversal symmetry.

In the other extreme, $c_g \gg c_0/\sqrt{\varepsilon_1 \mu_1}$, shown in Fig. 1C, we may still cling onto the Bloch wave picture except that the band gaps are regions of complex $\omega$ rather than of complex $k$. In these gaps waves can lose or gain energy, a process of parametric amplification [19].

The focus of our interest will be in the middle of these extremes: the luminal region in which the speed of the grating approaches that of light in the medium. Fig. 1D shows the catastrophe that occurs when $c_g = c_0/\sqrt{\varepsilon_1 \mu_1}$: all the forward travelling waves become degenerate. Whereas the band gaps formed between forward and backward travelling waves dominate scattering outside this regime, it is forward-forward scattering that dominates here. There is a clearly defined range of $c_g$ within which the band picture is invalid as already shown in earlier works [5]. In this range light is not scattered by the structure but is captured and localized, carried along with the structure with velocity $c_g$, amplified, and ejected from the medium as a series of pulses as shown in Fig. 2. Here a picture in terms of freely propagating waves is meaningless. Instead we look to the basic elements of the electromagnetic field: the lines of force embodied in the $D$ and $B$ fields. Because their velocity varies with position in the structure there will be a point towards which they migrate, an accumulation point [16,20], and here they are compressed and our new mechanism of amplification comes into play.

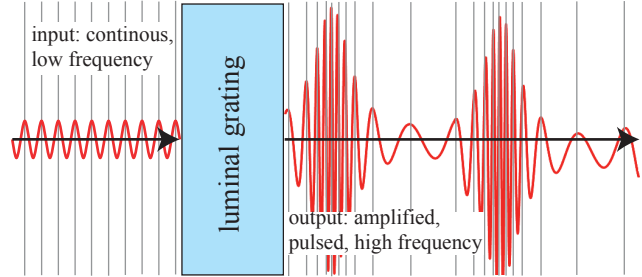

Fig. 2. Schematic figure of the effect of a luminal grating, travelling to the right, on a plane wave incident from the left. The grating has little effect on waves incident from the right. Note the compression of phase which mirrors the compression of lines of force shown schematically as grey lines.

The parameters $\varepsilon(x - c_g t)$ and $\mu(x - c_g t)$ define the impedance,

$$\mu(x - c_g t)/\varepsilon(x - c_g t) = Z^2, \qquad (3)$$

and refractive index,

$$\mu(x - c_g t)\varepsilon(x - c_g t) = n^2 \qquad (4)$$

Variations in the impedance are responsible for back scattering and hence for the band gaps. A constant $Z$ eliminates back scattering, removing all gaps so that there is no parametric amplification. Conversely the refractive index is responsible for forward scattering. Because of degeneracy of the forward travelling waves in a near luminal system forward scattering is of central importance and back scattering a distraction which we shall neglect. This is an exact statement if $Z$ is constant but approximately true if $\alpha_\varepsilon \ll 1, \alpha_\mu \ll 1$. Eliminating all band gaps and hence all parametric mechanisms allows us to identify our new mechanism in its purest form.

When back scattering is negligible the lines of force are conserved. Therefore any gain in energy can only come from compression of the lines into a sharp pulse. If their local density is increased by a factor of $f$, then the local energy density increases by $f^2$ and hence there is not only a local increase in energy but also a net increase.

Here we stress that this is an entirely novel insight into an amplification process. We have already identified parametric amplification which occurs when band gaps open and give rise to complex values of $\omega$ and wave fields that grow in time. No conservation is at work here, simply a uniform addition of lines of force. Next consider a slab of uniform gain medium characterized by a complex refractive index $n = n_r - in_i$ so that a wave of frequency $\omega$ injected into the system acquires a complex wave vector,

$$k = (n_r - in_i)\omega/c_0 \tag{5}$$

and as a result the wave amplitude increases exponentially with penetration into the medium. Cleary there is no conservation of lines of force here. Many more are ejected from the far side of the medium than enter.

## 2. Equation of motion in the absence of backscattering

We start from Maxwell's equations,

$$\nabla \times \mathbf{E} = -\frac{\partial \mathbf{B}}{\partial t} = -\frac{\partial \hat{\mu}\mu_0 \mathbf{H}}{\partial t}, \quad \nabla \times \mathbf{H} = +\frac{\partial \mathbf{D}}{\partial t} = +\frac{\partial \hat{\varepsilon}\varepsilon_0 \mathbf{E}}{\partial t} \tag{6}$$

where $\hat{\mu}$ and $\hat{\varepsilon}$ are operators which depend upon space and time and may be non local and dispersive, We impose the no back scattering condition,

$$\hat{\mu}\mu_0 = Z^2 \hat{\varepsilon}\varepsilon_0 \tag{7}$$

where $Z$ is the impedance of the medium and is a constant real number. If we assume that the fields depend only on $(x,t)$ then at normal incidence Maxwell's equations become,

$$\frac{1}{Z}\frac{\partial E_y}{\partial x} = -\frac{\partial \hat{\varepsilon}\varepsilon_0 (ZH_z)}{\partial t}, \quad \frac{1}{Z}\frac{\partial (ZH_z)}{\partial x} = -\frac{\partial \hat{\varepsilon}\varepsilon_0 E_y}{\partial t} \tag{8}$$

These equations are symmetric under exchange of $(E_y, ZH_z)$ and therefore solutions factorize into the symmetric and antisymmetric,

$$E_y = \pm ZH_z \tag{9}$$

The plus sign corresponds to forward travelling, and the minus sign to backward travelling waves obeying a first order partial differential equation,

$$+\frac{\partial D_y}{\partial t} = -\frac{\partial H_z}{\partial x} = \mp\frac{1}{Z}\frac{\partial E_y}{\partial x} \tag{10}$$

If we assume that $Z, \varepsilon$ are both real, and that $\varepsilon$ is independent of frequency and a local operator, then $D$ and $E$ have the same phase and there is a further factorization of (10) into real and imaginary parts,

$$+\frac{\partial |D_y|}{\partial t} = \mp\frac{1}{Z}\frac{\partial |E_y|}{\partial x}, \quad +|D_y|\frac{\partial \phi}{\partial t} = \mp\frac{1}{Z}|E_y|\frac{\partial \phi}{\partial x} \tag{11}$$

where $\phi$ is the phase of $D_y$ and, in consequence of (9), that of all four fields.

## 3. Conservation of lines of force

Suppose that,

$$\mu = Z^2\varepsilon = Z^2\varepsilon_1, \quad t < 0 \tag{12}$$

where $\varepsilon_1$ is independent of time. At $t=0$ we turn on the space-time dependence and ask how the total number of lines of force in the system changes,

$$\frac{\partial}{\partial t}\int_{-\infty}^{+\infty}|D_y|dx = \int_{-\infty}^{+\infty}\mp\frac{1}{Z}\frac{\partial |E_y|}{\partial x}dx = 0 \tag{13}$$

where we have substituted from (11). We assume that the fields have finite spatial extent and vanish at infinity which of course leaves the possibility of taking the limit of infinitely extended fields and so encompasses all situations considered here.

This proves the weak form of the theorem showing that the only way energy can be added to the system is by rearranging the lines of force.

If we make some further assumptions about $\varepsilon$ a stronger form of the theorem can be found. We now assume that $\varepsilon$ has the form,

$$\varepsilon(x - c_g t) \tag{14}$$

in other words it has a fixed profile moving with uniform velocity $c_g$. We work with new variables defined by,

$$X = x - c_g t, \quad t' = t \tag{15}$$

whereupon (11) becomes,

$$-c_g\frac{\partial |D_y|}{\partial X} + \frac{\partial |D_y|}{\partial t'} = \mp\frac{1}{Z}\frac{\partial |E_y|}{\partial X}, \quad -c_g\varepsilon\varepsilon_0\frac{\partial \phi}{\partial X} + \varepsilon\varepsilon_0\frac{\partial \phi}{\partial t'} = \mp\frac{1}{Z}\frac{\partial \phi}{\partial X} \tag{16}$$

The fields remain unchanged as they correspond to the original fields observed in the stationary frame. We ask if the number of field lines contained between two points,

$$x - c_g t = X_1, \quad x - c_g t = X_2 \tag{17}$$

is constant,

$$\frac{\partial}{\partial t}\int_{X_1}^{X_2}|D_y|dx = \int_{X_1}^{X_2}\left[\mp\frac{1}{Z}\frac{\partial|E_y|}{\partial X}+c_g\frac{\partial|D_y|}{\partial X}\right]dX = \left[\mp\frac{1}{Z}|E_y|+c_g|D_y|\right]_{X_1}^{X_2}$$
$$=\left[\left(\mp c_l(X)+c_g\right)|D_y|\right]_{X_1}^{X_2}$$
(18)

where $c_l(X) = c_0/(Z\varepsilon(X))$ is the local velocity of light. Thus the number of lines of force between $X_1, X_2$ is conserved if we are concerned with waves travelling in the same direction as the profile where the '−' sign applies in (18) and the two points specified meet the condition,

$$c_l(X_1) = c_l(X_2) = c_g \tag{19}$$

Hence the region between these points constitutes a trap in which, depending on the detailed variation of the profile, the lines of force may be continuously squeezed together and in the process energy pumped into the system. This local form of the theorem does not apply to waves travelling in the opposite direction to the profile where the '+' sign applies in (18) and they are not trapped.

## 4. Energy density

Remembering our assumption that $\hat{\mu}\mu_0 = Z^2\hat{\varepsilon}\varepsilon_0$ the energy density is given by,

$$U(X,t) = \varepsilon_0\varepsilon(X)|E_y(X,t)|^2 \tag{20}$$

and obeys the following equation,

$$\frac{\partial \ln U}{\partial t'} = \left[c_g - c_l(X)\right]\frac{\partial \ln U}{\partial X} + \left[c_g + c_l(X)\right]\frac{\partial \ln \varepsilon(X)}{\partial X} \tag{21}$$

which follows from (11). The second term on the rhs pumps energy into the system but equally important is the first term which derives from the Poynting vector. The system gains energy whenever the second term is positive, but energy is mobile if the local velocity of light is different from that of the grating: energy will flow away from the point of its creation. We discuss this flow below in the context of a periodic structure, and its role in producing highly focused pulses of radiation.

In the case of a periodic grating,

$$\varepsilon = \varepsilon_1\left[1+2\alpha\cos(gx-\Omega t)\right] \tag{22}$$

we can give some physical insight into energy growth.

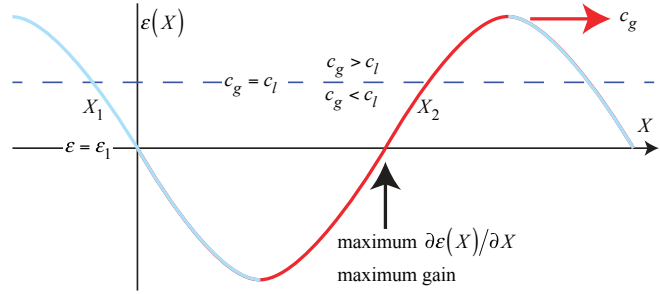

Fig. 3. Profile of the variation of $\varepsilon$ as a function of $X = x - c_g t$ shown oscillating about $\varepsilon = \varepsilon_1$. The red portion of the curve shows where time variation is pumping energy into the system, and the cyan portion where energy is being extracted. The dashed line indicates the value of $\varepsilon$ at which the local velocity is the same as that of the grating.

Fig. 3 shows the profile of the grating in the moving frame where $X_1, X_2$ are points where the local velocity of light equals that of the grating. As explained above, the last term in (21) is responsible for pumping energy into the system and is a maximum where the derivative of $\varepsilon$ is a maximum; in half the cycle this term is negative and takes energy from the system. The penultimate term in (21) arises from the Poynting vector and redistributes energy within the confines of $X_1 < X < X_2$. When the $c_g = c_l$ line coincides with the $X$ axis the two terms work together to give peak energy density at the point of maximum gain, otherwise the point of peak energy density drifts away from this point. Ultimately if the grating moves much quicker or much slower than any of the light within, no valid $X_1, X_2$ points can be found, the strong form of our theorem fails, and lines of force can now escape from one period of the grating to another, though still with overall conservation.

We can find an approximate but accurate solution to (21) for the case of the periodic grating. Here we give a derivation for the case of a weak modulation of $\varepsilon$ which approximately satisfies the no back scattering condition. Results for modulating both $\varepsilon$ and $\mu$ can easily be found by modifying (23) and (24).

Making the substitution,

$$\tau = t' \frac{c_0}{\sqrt{\varepsilon_1}} \qquad (23)$$

and assuming that $\alpha$ is small (21) can be written,

$$\frac{\partial L}{\partial \tau} = -g(2+\delta)\alpha \sin(gX) + [\delta + \alpha \cos(gX)]\frac{\partial L}{\partial X} \qquad (24)$$

where,

$$L = \ln U', \quad c_g = (1+\delta)c_0/\varepsilon_1 \qquad (25)$$

we solve (24) by successive approximations. To zeroth order we neglect the second term on the rhs of (24),

$$L_0 = -g(2+\delta)\alpha \sin(gX)\tau \qquad (26)$$

This implies that the weight of the gain occurs in the close vicinity of,

$$\sin(gX) = -1, \quad gX = 3\pi/2, \quad \cos(gX) \approx (gX - 3\pi/2) \qquad (27)$$

To calculate the first order we substitute the zeroth order into the missing term,

$$\frac{\partial L_1}{\partial \tau} = -g(2+\delta)\alpha \sin(gX) + [\delta + \alpha \cos(gX)]\frac{\partial L_0}{\partial X}$$
$$= -g(2+\delta)\alpha \sin(gX) - [\delta + \alpha \cos(gX)]\cos(gX)g(2+\delta)\alpha\tau \quad (28)$$

Hence,

$$L_1 = L_0 + \Delta_1 = -g(2+\delta)\alpha \sin(gX)\tau$$
$$-[\delta + \alpha \cos(gX)]\cos(gX)g(2+\delta)\alpha g\frac{\tau^2}{2} \quad (29)$$

Proceeding in this manner we find,

$$\Delta_n = -[\delta + \alpha \cos(gX)]^2 \frac{1}{2\alpha^2}(2\alpha g)^{n+1}\frac{\tau^{n+1}}{(n+1)!}$$
$$+\delta[\delta + \alpha \cos(gX)]\frac{2}{\alpha^2}(\alpha g)^{n+1}\frac{\tau^{n+1}}{(n+1)!} \quad (30)$$

Summing the terms to infinity,

$$L = -g(2+\delta)\alpha \sin(gX)\tau + \sum_{n=1}^{\infty}\Delta_n$$
$$= -g(2+\delta)\alpha \sin(gX)\tau$$
$$-[\delta + \alpha \cos(gX)]^2 \frac{1}{2\alpha^2}\left[-1 - 2\alpha g\tau + e^{2\alpha g\tau}\right]$$
$$+\delta[\delta + \alpha \cos(gX)]\frac{2}{\alpha^2}\left[-1 - \alpha g\tau + e^{\alpha g\tau}\right] \quad (31)$$

and,

$$U(X,t) = \exp\left\{\begin{array}{l}-g(2+\delta)\alpha\sin(gX)\tau \\ -[\delta+\alpha\cos(gX)]^2\frac{1}{2\alpha^2}\left[-1-2\alpha g\tau+e^{2\alpha g\tau}\right] \\ +\delta[\delta+\alpha\cos(gX)]\frac{2}{\alpha^2}\left[-1-\alpha g\tau+e^{\alpha g\tau}\right]\end{array}\right\} \quad (32)$$

Note that the expression is an exact solution of (24) if $\delta = 0$, $gX = 0, 3\pi/2$. The expression describes formation of a pulse contained within a period of the grating whose position shifts as the grating velocity deviates from the average speed of light as described by $\delta$. The result is independent of the frequency of incident radiation, $\omega$, and the modulation frequency, $\Omega$, appears only through $g = \Omega/c_g$ as a scaling variable for length.

The first term in brackets in Eq. (32) arises from the rate of change of $\varepsilon$ and is a maximum when that rate is a maximum. It gives rise to exponential growth in amplitude. Also there is a point in the grating where $\sin(gX) = 0$ and gain switches to loss. The other terms arise from the flow of energy and substantially change the shape of the pulse that forms and where the pulse forms. Narrowing of the pulses is also exponential as our theory requires to be the case.

If the grating travels at the average velocity of light, $\delta = 0$, our approximations are rather accurate and in fact are exact at $gX = \pi/2$ and at $gX = 3\pi/2$, the maximum loss and maximum gain points in the medium, where only the first term matters. When $\delta \neq 0$ energy drifts away from the point of maximum creation towards the point where the velocity of light is the same as that of the grating. Peak energy density lies between the two. We give a further discussion of accuracy later in the paper when comparing to transfer matrix simulations.

Here is the origin of trapping lines of force to which we alluded earlier. If there is no point within the grating where the local velocity of light is the same as that of the grating, then lines of force escape into the next period and the growth mechanism collapses and we revert to a Bloch wave description. The condition for the growth mechanism to operate is [2],

$$1/\sqrt{1+2\alpha} < \sqrt{\varepsilon_1}\, c_g/c_0 < 1/\sqrt{1-2\alpha} \tag{33}$$

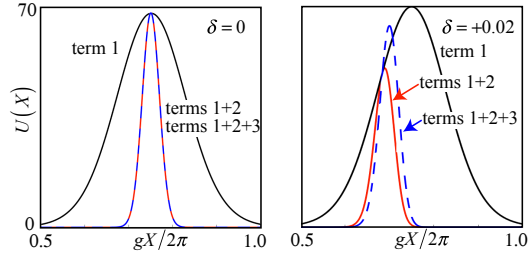

Fig. 4. Evaluation of (32) for transmission through a finite luminal medium for $\delta = 0$ (left) and $\delta = +0.02$ (right), thickness $\tau = t'c_0/\sqrt{\varepsilon_1} = 600$, and $\alpha = 0.05$, at the exit surface, as a function of $gX$. In each case we include successively the first term in (32): Outer curve; plus the second: full inner curve, and plus the second and third: dashed inner curve.

In Fig. 4 we show the effect of the three contributions to $U(X,t)$. When $c_g = c_0/\sqrt{\varepsilon_1}$, $\delta = 0$, energy accumulates at the point of maximum growth and the role of the second term is to sharpen the peak as energy migrates inwards. The third term makes no contribution in this case. In contrast when the grating is moving faster, $c_g > c_0/\sqrt{\varepsilon_1}$, $\delta = 0.02$, the third and second terms play a role both in sharpening the peak and in moving it back along the grating.

We recognize that (32) implies the following density of lines of force

$$|D(X,t)| = \sqrt{\varepsilon\varepsilon_0 U(X,t)}$$
$$\approx \sqrt{\varepsilon\varepsilon_0} \exp\left\{ \begin{array}{l} -\dfrac{g}{2}(2+\delta)\alpha\sin(gX)\tau \\ -[\delta+\alpha\cos(gX)]^2 \dfrac{1}{4\alpha^2}\left[-1-2\alpha g\tau + e^{2\alpha g\tau}\right] \\ +\delta[\delta+\alpha\cos(gX)]\dfrac{1}{\alpha^2}\left[-1-\alpha g\tau + e^{\alpha g\tau}\right] \end{array} \right\} \tag{34}$$

This is a very curious time evolution containing a double exponential, the function of which is to narrow the width of the pulse and ensure conformity to the flux conservation law.

## 5. Dependence of the phase on time

From (16), (22) and (23),

$$\frac{\partial \phi}{\partial t'} = +c_g \frac{\partial \phi}{\partial X} - \frac{c_0}{\sqrt{\varepsilon}} \frac{\partial \phi}{\partial X} = +(1+\delta)\frac{c_0}{\sqrt{\varepsilon_1}} \frac{\partial \phi}{\partial X} - \frac{c_0}{\sqrt{\varepsilon}} \frac{\partial \phi}{\partial X}$$
$$\approx +\delta \frac{c_0}{\sqrt{\varepsilon_1}} \frac{\partial \phi}{\partial X} + \alpha \cos(gX) \frac{c_0}{\sqrt{\varepsilon_1}} \frac{\partial \phi}{\partial X}$$
(35)

and hence,

$$\frac{\partial \ln \psi}{\partial \tau} = -\alpha g \sin(gX) + \left[\delta + \alpha \cos(gX)\right] \frac{\partial \ln \psi}{\partial X}$$
(36)

where,

$$\partial \phi / \partial X = \psi$$
(37)

We can retrace the steps taken in deriving $U$. We concentrate not on the maximum rate of growth of $\phi$, rather that of $\psi$ and find to first order,

$$L'_0 = -\alpha g \sin(gX) \tau$$
(38)

Back substituting and making successive approximations,

$$\Delta L'_n = -\left[\delta + \alpha \cos(gX)\right] \alpha^n g^{n+1} \cos(gX) \frac{\tau^{n+1}}{(n+1)!}$$
$$+ \left[\delta + \alpha \cos(gX)\right]^2 \alpha^{n-1} g^{n+1} \frac{\tau^{n+1}}{(n+1)!}$$
$$- \left[\delta + \alpha \cos(gX)\right]^2 2^{n-1} \alpha^{n-1} g^{n+1} \frac{\tau^{n+1}}{(n+1)!}$$
(39)

following through to,

$$\frac{\partial \phi}{\partial X} = \psi \approx k_1 \exp \left\{ \begin{array}{l} -g\alpha \sin(gX)\tau \\ -\left[\delta+\alpha\cos(gX)\right]\alpha^{-1}\cos(gX)\left(-1-\alpha g\tau + e^{\alpha g \tau}\right) \\ +\left[\delta+\alpha\cos(gX)\right]^2 \alpha^{-2}\left(-\frac{3}{4}-\frac{1}{2}\alpha g\tau + e^{\alpha g\tau} - \frac{1}{4}e^{2\alpha g\tau}\right) \end{array} \right\}$$
(40)

The prefactor, $k_1 = k_0/\sqrt{\varepsilon_1}$, is the wave vector in the background medium and gives the rate of change of $\phi$ at $\tau = 0$. Compare this equation to our expression for the flux density, (34), which to the accuracy of our approximations shows that the phase is compressed in the same manner as the lines of force.

Integrating (40) we find the phase,

$$\phi(X,\tau) \approx k_1 e^{-g\alpha \sin(gX)\tau} \frac{\sqrt{\pi}}{2g} b(\tau)\left[1+\text{erf}(x_0(\tau))\right]$$
(41)

where,

$$b(\tau) = \frac{1}{\sqrt{-\frac{1}{4} - \frac{1}{2}\alpha g \tau + \frac{1}{4}e^{2\alpha g \tau}}}, \quad x_0(\tau) = \frac{(gX - 3\pi/2)}{b(\tau)} \qquad (42)$$

The phase depends on the modulation frequency $g = \Omega/c_g$ as a scaling variable for the length however the number of oscillations in one cycle is given by $k_1/g$ and so does depend on the input frequency.

## 6. Comparing theory to transfer matrix calculations

In Fig. 5 we compare the analytic result with transfer matrix based simulations [16]. The first two figures show transmission through slabs of grating of two different thicknesses whilst the grating travels at the average velocity of light, $\delta = 0$. The choice the input frequency, $\omega = 1$, does not affect transmission intensities but $\Omega = 0.07$ the modulation frequency dictates the number of oscillations of phase per cycle. The choice of a small ratio $\Omega/\omega$ both serves to provide a rich population of phase oscillations and to demonstrate that modulations can be effective whilst of a much lower frequency than the waves acted upon.

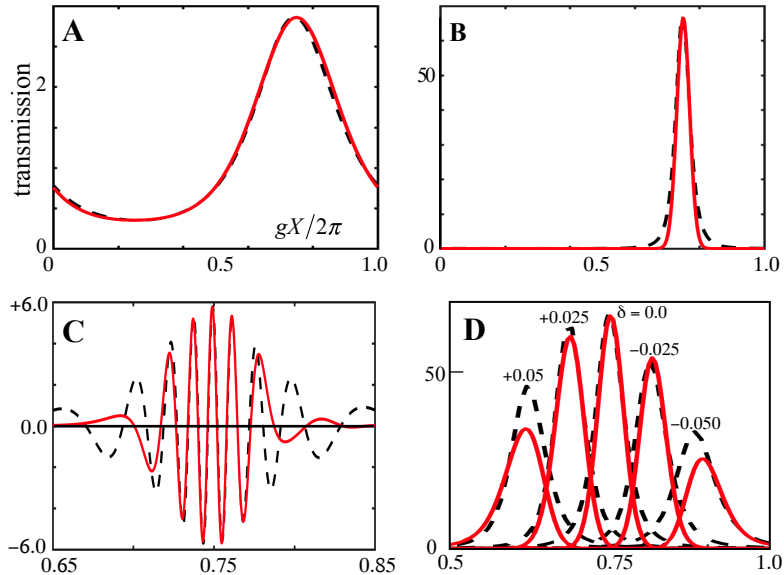

Fig. 5. Intensity transmitted through a finite luminal medium, $\delta = 0$, at the exit surface, as a function of $gX$, **A**: after a short time propagating along the medium, $\alpha = 0.05$, $\Omega = 0.07$, $\tau = 150$ and **B**: after a longer time, $\tau = 600$. The dashed black line is calculated from the transfer matrix, and the red line from the present approximate theory. **C**: same parameters as for **B** but here we plot the real part of the amplitude which includes a calculation of the phase where the incident wave has frequency $\omega = 1$. **D**: same parameters as for **B** except for $\delta \neq 0$.

In Fig. 5A the light has been in the grating for a short time, $\tau = 150$, in Fig. 5B for a longer time, $\tau = 600$ resulting in a higher narrower pulse. Our compression of lines of force theory predicts that the pulse height scales inversely as the square of the pulse width, which it does. Note the agreement with the transfer matrix calculations. In Fig. 5C we add the phase calculated analytically and compare the real part of the electric field with that calculated using the transfer matrix. Derivation of the phase formula follows closely that for the field amplitude. The formula is accurate where the linearisation assumption holds good.

Finally in Fig. 5D the calculation in 5B is extended to other grating velocities. Here the analytic result is less accurate (it depends on linearisation of $\varepsilon(X)$ about $gX = 3\pi/2$). The amplification mechanism is on the point of breakdown at $\delta = \pm 0.05$ beyond which values a Bloch wave picture reasserts itself. Note how the peak lags behind the point of maximum gain when $\delta > 0$ and conversely when $\delta < 0$.

## 7. Discussion

Our idealized model which assumes no loss and no dispersion will inevitably be compromised to some extent by the materials available to us.

First let us consider the problem of loss. Adjusting (10) to include uniform loss,

$$+\frac{\partial D_y}{\partial t} = +\frac{\partial (\varepsilon_r + i\varepsilon_i) E_y}{\partial t} = +\varepsilon_r \frac{\partial E_y}{\partial t} + i\varepsilon_i \frac{\partial E_y}{\partial t} + E_y \frac{\partial \varepsilon_r}{\partial t} = \mp \frac{1}{Z}\frac{\partial E_y}{\partial x} \qquad (43)$$

we see that the loss term represented by $\varepsilon_i$ competes directly with the time dependence of $\varepsilon_r$. Within the luminal region $\partial \varepsilon_r/\partial t$ results in excitations of higher frequencies of the form $\omega + n\Omega$ which add coherently to form the pulse. Clearly from (43) the higher frequencies will be more susceptible to loss than the lower ones and we can expect a cut off in the spectrum approximately when,

$$\varepsilon_i(\omega + n_{cut}\Omega) \approx \frac{\partial \varepsilon_r}{\partial t} \qquad (44)$$

and the pulse saturates at a finite width. In addition loss will result in attrition of the number of lines of force which are no longer conserved. The gain process as a whole will cease for higher values of loss,

$$\varepsilon_i(\omega + \Omega) \approx \frac{\partial \varepsilon_r}{\partial t} \qquad (45)$$

which implies a minimum value of $\partial \varepsilon_r/\partial t$ for these effects to materialize.

Another consideration is the practicality of modulating material properties at very high frequencies. At RF where a rich variety of material properties is available there will be no problem and several schemes involving varactors have been proposed [13] but at higher frequencies only weak modulation can be expected. However there is a let out clause in that only the *speed* of the modulation need keep pace with the radiation, the modulation *frequency*, $\Omega$, can in principle be much lower provided that $c_g = \Omega/g$ is fast enough. In lossy systems limits are imposed by (45) because a small value of $\Omega$ implies a low rate of modulation.

Choosing a value of $\Omega \ll \omega$ helps sidestep the issue of dispersion. When $\varepsilon$ is a function of frequency each frequency excited has a different velocity and therefore a different accumulation point within the cycle, see Fig. 3, and as a result does not contribute to give a perfectly coherent peak. In fact if dispersion is so severe that some frequencies escape the trap entirely they no longer contribute to amplification. This has been demonstrated computationally in an earlier publication [16] by means of transfer matrix calculations. Most materials have a range of frequencies over which dispersion is small so that by choosing $\Omega \ll \omega$ the system can stay within the gain regime to all practical purposes providing that losses are modest.

Apart from RF systems graphene is a promising candidate for THz frequencies. It is known that the conductivity of graphene can be modulated at rates exceeding 100GHz via both electro-optic [21] and all-optical mechanisms [22] and the THz surface plasmons of graphene may be the first excitations to be amplified in this fashion. We have previously suggested double layer graphene as a possible candidate since the linearity of the acoustic plasmon mode in this configuration enables the circumvention of the aforementioned dispersion effects [16].

Seemingly related but well-distinct light-amplification mechanisms have recently been proposed, which exploit the use of DC currents in graphene, where the velocity of the carriers can be uncommonly high, and the mechanism has been associated with the phenomenon of Landau damping [23], suggesting a possible common underlying origin, although no charge motion is present in our case.

Although our discussion has been entirely in terms of electromagnetic waves, similar processes will apply to other waves: to water waves and particularly to acoustic waves. In many ways acoustic systems could be much more amenable to realization: frequencies are much lower removing the problem of modulation speed and introducing the possibility of electronic control [24]. Also many acoustic systems have extremely low loss and low dispersion alleviating other difficulties.

## 8. Acknowledgments

We thank the following for support: P.A.H. acknowledges funding from Fundação para a Ciencia e a Tecnología and Instituto de Telecomunicaçõoes under Projects CEECIND/03866/2017 and UID/EEA/50008/2020. E.G. is supported through a studentship in the Center for Doctoral Training on Theory and Simulation of Materials at Imperial College London funded by the Engineering and Physical Sciences Research Council (EPSRC) grant number EP/L015579/1 and by EPSRC grant number EP/T51780X/1. J.B.P. acknowledges funding from the Gordon and Betty Moore Foundation.

## Disclosures

The authors declare no conflicts of interest.